\documentclass[12pt,manuscript]{aastex}

\received{}
\revised{}
\accepted{}

\shortauthors{Jones et al.}
\shorttitle{N159 in the LMC}

\begin{document}

\title{{\it Spitzer} IRAC Observations of Star Formation in N159 in the LMC}
\author{Terry J. Jones, Charles E. Woodward, Martha L. Boyer, 
\\ Robert D. Gehrz, and Elisha Polomski}
\affil{Department of Astronomy, University of Minnesota, 116 Church 
Street S.E., \\ Minneapolis, MN 55455 \\ tjj, chelsea, mboyer, gehrz, 
elwood@astro.umn.edu}

\begin{abstract}

We present observations of the giant HII region complex N159 in the LMC using IRAC on the {\it Spitzer Space Telescope}. One of the two objects previously identified as protostars in N159 has an SED consistent with classification as a Class I young stellar object (YSO) and the other is probably a Class I YSO as well, making these two stars the 
youngest stars known outside the Milky Way. We identify two other sources that may also be Class I YSOs. One component, N159AN, is completely hidden at optical wavelengths, but is very prominent in the infrared. The integrated luminosity of the entire complex is L $\approx 9\times10^6$~L$_{\odot}$, consistent with the observed radio emission assuming a normal Galactic initial mass function (IMF). There is no evidence for a red supergiant population indicative of an older burst of star formation. The N159 complex is 50~pc in diameter, larger in physical size than typical HII regions in the Milky Way with comparable luminosity. We argue that all of the individual components are related in their star formation history. The morphology of the region is consistent with a wind blown bubble $\approx 1-2$~Myr-old that has initiated star formation now taking place at the rim. Other than its large physical size, star formation in N159 appears to be indistinguishable from star formation in the Milky Way. 

\end{abstract}

\keywords{stars: formation --- (ISM:) HII regions --- (galaxies:) 
Magellanic Clouds} 

\section{INTRODUCTION}

The Magellanic Clouds, the nearest moderate sized galaxies to our own 
Milky Way, fortuitously lie in a direction well out of the plane and are 
relatively free of foreground extinction. As a result, they present a unique opportunity to study the process of star formation in a galaxy other than our own. N159 is a group of HII regions cataloged by \citet{hen56} who gave separate letters to distinguish between the clumps of nebulosity found within the complex. The first extragalactic protostar was discovered in the N159 complex \citep{gat81}, as well as the the first extragalactic Type~I~OH maser 
\citep{cas81}, and the region contains an H$_{2}$O maser \citep{sca81}.

Radio observations of the LMC with the Molonglo Synthesis 
Telescope \citep[MOST,][]{mil84} revealed high surface brightness, optically 
thin, thermal radio emission characteristic of HII regions powered by 
ionizing photons from OB stars. A higher resolution radio map by 
\citet{hun94} confirm the morphology seen in the MOST maps. CO observations 
\citep{coh88, bol00, joh98} reveal a very high concentration of molecular 
gas at the location of N159. 

\citet{gat81} searched N159 at broadband {\it K} (2.2~\micron ) looking for 
very red objects that might be the analog of sources such as BN in the 
Milky Way. They found such an object, with $K-L^\prime = +2.7$~mag color 
and no detectable Br$\gamma$ emission at 2.16~$\micron$, which they 
suggested was a protostar. Subsequently, \citet{gat82} found a candidate 
protostellar object in the Small Magellanic Cloud (SMC) while \citet{jon86} 
found a second candidate in the N159 complex. Our understanding of YSOs has 
advanced considerably since these discoveries (for a discussion of low mass YSOs, see \citet{ada87, shu87} and for high mass YSOs, see \citet{chu02, gar99}).
Currently YSOs are classified in an evolutionary sequence from Class 0 to 
III. Class 0 stars, the youngest, show no near infrared (near-IR) 
counterpart and are detected primarily at sub-MM wavelengths. Such objects 
will be impossible to find in N159 at present because of beam confusion 
with current far-IR and sub-mm facilities. Class I YSOs, however, typically 
peak in the mid-IR and have detectable near-IR flux \citep{shu87, whi03}. A 
Class I YSO may, or may not, depending on mass and evolutionary stage, be 
producing ionizing radiation \citep{chu02}. 

The first far-IR detection of N159 was made with the Air Force Geophysics 
Laboratory (AFGL) survey \citep{pri76} and a single beam observation from 
the {\it Kuiper Airborne Observatory} (KAO) was made by \citet{wer78}. 
\citet{jon86} mapped several HII regions, including N159, using the KAO at 
40\arcsec \ resolution and found very close correspondence between the far-IR 
and MOST maps, as expected for dusty HII regions. H$\alpha$ observations of 
N159 have revealed locations with high surface brightness that were studied by 
\citet{hey82} in N159A and by \citet{hey99} in N159-5 (The Papillon). The 
stellar content associated with N159A was studied at optical wavelengths by 
\citet{deh92} while the stellar content around the Papillon was studied at 
near-IR ({\it JHK}) wavelengths by \citet{mey04}. 

There are three questions regarding star formation in N159 we wish to explore 
using Infrared Array Camera \citep[IRAC,][]{fazio04} images 
obtained with the {\it Spitzer Space Telescope} \citep[{\it Spitzer},][]{wer04}. 
First is the nature of the two protostars discovered by \citet{gat81} and 
\citet{jon86}. In light of our new measurements and a more 
contemporary understanding of very young stars, how should the two 
protostars in N159 be classified? Is there any evidence for additional protostars missed in the earlier near-IR surveys?

A second question we wish to address is the nature of the upper end of the 
Initial Mass Function (IMF) in the LMC. Based on far-infrared (far-IR) 
observations of giant LMC HII regions, \citet{jon86} suggested N159 was 
producing more ionizing photons for its luminosity than it should, given 
their estimate of the total luminosity. Perhaps the IMF was weighted to 
producing more very massive stars than is the case in the Milky Way. 
Subsequent studies in the optical, in particular \citet{mas98}, argue that 
the LMC has a normal IMF at the extreme high mass end. Our {\it Spitzer}
observations fill in a portion of the spectral energy distribution (SED) 
missing from the earlier study by \citet{jon86} enabling a more accurate 
assessment of N159's luminosity.

Thirdly, we seek to understand the morphology of N159, which is larger in physical size ($\approx 50$~pc) than typical for Galactic HII regions. Since there is a significant reservoir of molecular gas in the vicinity of N159, the individual components of the complex may be unrelated and simply the product of randomly initiated star formation at locations of dense molecular gas. Alternatively, the morphology and stellar content may suggest a scenario where the star formation within N159 was triggered by a single, relatively recent event less than 2~Myr ago. 

In this paper we present {\it Spitzer} images of N159 using the four channels 
of IRAC with band centers at $3.5, \ 4.5, \ 5.7,$ and $7.9~\micron$. 
In \S2 we describe the observations and data reduction. In \S3 we examine the 
general morphology of the nebula and make comparisons to observations at 
other wavelengths. In \S4 we study the point sources seen in 
the IRAC maps, including the two protostars. Finally, we discuss the global 
properties and evolutionary state of the N159 complex and present a 
scenario for wind driven star formation on 50~pc scales. 

\section{OBSERVATIONS AND DATA REDUCTION \label{obs}}

Observations were made using all four bands of IRAC on {\it Spitzer}
as part of a Guaranteed Time Observing Program (Program ID 124) conducted by
{\it Spitzer} Science Working Group member R.~D. Gehrz. The 
band centers are at 3.548, 4.492, 5.661 and
7.87~\micron. The adopted zero magnitude fluxes were 277.9~Jy (band~1), 
179.5~Jy (band~2), 116.6~Jy (band~3) and 63.1~Jy (band~4) as described
in Table 5.1 of the IRAC Data 
Handbook\footnote{http://ssc.spitzer.caltech.edu/irac/dh/iracdatahandbook.html}
V1.0. The IRAC images contain an off-target field of comparable area on 
the sky and we were able to make use of this extra data for 
computing star counts in band~1 (see \S4.2).

The raw {\it Spitzer} data were processed with version 9.5.0 of the 
Spitzer Science Center (SSC) pipeline. Post-BCD processing was carried out 
using the 2004 June 6 Linux version of the SSC Legacy MOPEX software. Four 
steps of MOPEX were implemented: Cosmetic Fix, Background Matching, 
Mosaicker, and Astronomical Point Source Extraction (APEX). The Cosmetic Fix 
eliminated column pull-down and mux-bleed artifacts associated with 
3.5~\micron \  and 4.5~\micron \ data. Background matching was 
performed by minimizing the pixel differences in overlapping areas 
with respect to a constant offset computed by the program. Cosmic rays 
and other outliers were detected and eliminated with Mosaicker. The 
cosmetic, background matching, and cosmic ray corrections were all applied 
to mosaics created for each channel by the Mosaicker. APEX detected point 
sources in each basic calibrated data (BCD) image and computed 
point-spread functions (PSFs) and aperture fluxes for each source. The 
PSFs were created in-flight using observations of a bright star taken 
during in-orbit checkout (IOC) of the spacecraft.

The N159 region contains considerable nebulosity and this often caused 
confusion in APEX for fainter sources. We dealt with this problem by 
visually examining the band~1 mosaic image and eliminating any 
APEX source that was not clearly a point source. We also set a cutoff in 
our point source data set at [3.5] = 16.0~mag (corresponding to 110.6~$\mu$Jy) 
for this study. We made a check of the APEX results by directly 
measuring the PSF for bright sources in all four bands and comparing the 
area under our PSF fit with the APEX values. Agreement was better than 
3\%~rms for bands~1 and 2 and about 5\% for bands~3 and 4. There were 
very few point sources in bands~3 and 4 since the spectral energy 
distribution (SED) for stars is falling rapidly at these longer
wavelengths. Several objects returned by APEX as point sources 
were actually slightly extended in one or more bands and these are 
discussed in the text (\S\ref{hii}). 

Table~\ref{master} contains the IRAC fluxes for all point sources returned by APEX for the N159 region subject to the restrictions described above. Column~[1] is a running number we will use in this paper to identify sources. Columns [2] and [3] are the J(2000) RA and Decl. in decimal degrees from the IRAC astrometry for band 1. Column~[4] is the flux reported by APEX in band 1 in units of $\mu$Jy. Column~[5] is the error on this flux reported by APEX. Columns [6]-[11] are the fluxes and errors for bands 2, 3, and 4. Point sources in the off-source portions of our IRAC images are not included in Table~\ref{master}.

Table~\ref{spsc} presents a subset of the entire point source data set 
that includes photometry from the {\it Two Micron All Sky Survey}\ (2MASS)
and elsewhere in the literature. The sources in Table~\ref{spsc}
were chosen either because they were bright ([3.5] $\leq 12$~mag), 
had been observed in some other wavelength regime, or were found to 
have interesting colors. Column~[1] is the running number from column~[1] of Table~\ref{master}. Column~[2] gives the 2MASS identification, which contains the J2000 
coordinates derived from 2MASS astrometry. Source No.~134 has no 
2MASS counterpart, and its {\it Spitzer} designation in col~[3]
contains the J2000 coordinates derived from IRAC astrometry. Agreement 
between IRAC and 2MASS coordinates was better than $0\farcs5$ rms. 
Column~[3] contains other identifications, 
including {\it Midcourse Space Experiment} (MSX) point 
sources, when found. Columns~[4] -- [11] list the photometry and 
column~[12] contains additional information on each source. Note 
that many band~3 and band~4 point sources were impossible to extract 
from the surrounding nebulosity with any certainty. Those sources in 
Table~\ref{spsc} with band~3 and 4 photometry were either very bright, 
or isolated from nebulosity.

\section{GENERAL MORPHOLOGY \label{genmorph}}

At optical wavelengths the N159 complex has a roughly round shape that 
spans a physical size of about 50~pc, assuming a distance modulus of 
18.5 to the LMC. Figure~\ref{tjj-f1} presents a gray scale image in IRAC 
band 1 (3.5~\micron) with a number of important objects identified that we 
will discuss later. The B5~Iab supergiant (Sk-69~254) is the visually 
brightest star in the region. The high surface brightness HII regions (the 
Papillon, N159A, and N159AN) and the two protostars (P1 and P2) are the 
brightest objects in the IRAC bandpasses. 

A false color image of N159 combining IRAC bands 1 (blue), 2 (green) and 4 
(red) is shown in Fig.~\ref{tjj-f2}. We have 
chosen the color balance so that all unreddened stars will appear 
blue in Fig.~\ref{tjj-f2}, even red giants. 
Stars later in spectral type than G0 develop photospheric CO absorption at 
4.6~$\micron$ and this causes the IRAC [3.5]-[4.5] color to become 
negative, hence the blue color \citep[see] []{rei02}. By band 4 
(7.9~\micron), most of the stars are very faint in comparison to the 
nebulosity, so there is little red component to their colors in this false 
color image. The brightest objects (e.g., P1) are set to white, and 
they are generally bright at all wavelengths. The nebulosity appears red or 
green, and is dominated by hot dust emission in band~2, and hot dust + 
polycyclic aromatic hydrocarbon \citep[PAH,][]{allam99} emission in band~4. 
We consider it unlikely that Br$\alpha$ emission can make a significant 
contribution to band 2. Examination of the transmission curve for 
band 2 and investigation 
of {\it Infrared Satellite Observatory} (ISO) spectra of dusty galactic HII 
regions such as W51~IRS2 \citep{slo03}, indicates that the Br$\alpha$ 
emission line has too small an equivalent width to contribute significantly 
to the total flux in band 2. There could be some contribution from PAH 
emission at 3.28~$\micron$ in band 1.

The 5~GHz contour map \citep{hun94}, overlying a gray scale IRAC 
band~1 image (Fig.~\ref{tjj-f1}), is shown in Fig.~\ref{tjj-f3}a. \citet{hun94} 
also present an H$\alpha$ contour map of the region and this is shown in 
Fig.~\ref{tjj-f3}b, also overlying a gray scale band 1 image
(Fig.~\ref{tjj-f1}). For clarity we have removed the inner contours from 
the 5 GHz map for the Papillon and N159AN and from the H$\alpha$ map for 
the Papillon for clarity. The three 
strongest 5~GHz radio peaks correspond to the Papillon, N159A and N159AN. 
N159A breaks up into two peaks with the western peak corresponding to an 
MSX point source G280.2342-31.5130. The eastern peak is close to, but is not coincident with P2. N159AN corresponds to the MSX point source 
G280.2203-31.5163 and the Papillon corresponds to the MSX point source 
G280.2128-31.4949. 

There is also a radio peak near the center of the nebula at RA = $5^{h} 
39^{m} 53^{s}$, Decl. = $-69\degr 45\arcmin 18\arcsec$ (J2000), northwest 
of the protostar P1 that shows up more clearly in the lower resolution MOST 
map \citep{mil84}. The radio peak is also listed in the MSX point source 
catalog as G280.2128-31.4949. We will discuss this source more in 
\S\ref{hii}. Note that N159AN does not appear in the H$\alpha$ map. 
\citet{hun94} concluded that this HII region must be behind significant 
interstellar extinction. We estimate an extinction of at least 4 magnitudes 
at V in order to completely hide the H$\alpha$ emission. 

The protostar P1 does not lie at any 5~GHz or H$\alpha$ peak and is not in 
the MSX point source catalog. P2 is on the NE edge of the eastern peak in 
the 5~GHz and H$\alpha$ distribution for N159A, but is not coincident with these peaks, nor is P2 in the 
MSX point source catalog. The Papillon, on the other hand, is very bright 
in H$\alpha$ \citep{hey99}, radio emission, and is only slightly extended 
in our IRAC images. N159AN also contains a very compact source of radio 
emission coincident with a prominent, slightly extended IR source. The 
western peak of N159A is coincident with a peak in the 5~GHz radio map and 
is associated with H$\alpha$ emission. We will argue in \S\ref{protos} that 
neither protostar is producing significant ionizing radiation, in contrast to these 
other point-like sources that are clearly associated with compact HII 
regions. 

Figure~\ref{tjj-f4} shows a contour map of CO emission from \citet{joh98} 
overlying the IRAC band~1 gray scale (Fig.~\ref{tjj-f1}). There are two 
prominent peaks in the CO distribution that are extensively discussed by 
\citet{bol00}. One is just east of the Papillon, and given the very bright 
H$\alpha$ from this source, the molecular gas probably lies behind or off 
to the side of this object. The western concentration is nearly coincident 
with N159AN, the completely obscured HII region. Using the H$_{2}$ column 
depth of $1.2 \times 10^{22}$~cm$^{-2}$  from \citet{bol00}, 
we would estimate A$_{V} 
= 3$~mag \citep[see also][]{mey04}. This $A_{V}$ value is less than our 
previous estimate of at least A$_{V} = 4$~mag of extinction toward this 
region. The discrepancy may be attributable to beam dilution in the CO 
observations. N159AN very likely lies behind or within a large, dense 
molecular cloud corresponding to the western peak in the CO distribution. 
The center of N159, which shows a local peak in the radio emission, shows 
much less CO emission than near the Papillon, N159A and at N159AN. 

\section{STELLAR CONTENT \label{stellcnt}}

\subsection{Color-Magnitude Diagram \label{cmd}}

There are several sources of photometry of point sources in N159 in 
addition to our new {\it Spitzer} observations 
presented in Table~\ref{spsc}. The 2MASS catalog contains photometry for the 
majority of the stars we detected at 3.5~\micron , including the two 
protostars. Higher spatial resolution {\it JHK} photometry of stars near 
the Papillon has been reported by \citet{mey04}, which also includes 
observations of P1. \citet{deh92b} tabulate deep optical CCD photometry of 
the region around N159A which they analyze in \citet{deh92}. These 
observations provide an excellent data base for comparison of photometry at 
optical, near-IR and mid-IR wavelengths at arcsecond spatial resolution. 

Figure~\ref{tjj-f5} is a color-magnitude diagram of the point sources in 
the N159 region using the [3.5] and [4.5] magnitudes from the IRAC images. 
Unlike more traditional color-magnitude diagrams such as an HR diagram, 
redder {\it Spitzer} colors do not necessarily correspond to cooler effective 
temperature. All MK spectral types earlier than G0 will have [3.5]-[4.5] 
colors very close to zero. Stars later than G0 begin to develop CO 
absorption in their atmospheres at 4.6~$\micron$ and consequently have 
[3.5]-[4.5] colors that become negative. This trend is indicated by the 
spectral type labels on the bottom axis stretching from -0.25 to 0.0 in 
[3.5]-[4.5] color. These fiducial points were determined by integrating ISO 
spectra \citep{slo03} of late type stars over the IRAC bandpasses in 
combination with the CO equivalent widths for late type stars determined by 
\citet{her02} (also based on ISO spectra). The majority of the point 
sources have colors in this range. Oxygen rich stars later than M5 in 
spectral type develop significant water vapor absorption in their 
atmospheres and we have not investigated this influence on [3.5]-[4.5] 
color. Also, we have not investigated the expected colors for carbon stars. 

The reddening vector is derived from the reddening law tabulated in 
\citet{rie85}. Note that significant reddening is required for otherwise 
normal stars to be shifted down and to the right in Fig.~\ref{tjj-f5}. Star 
No.~134, for example, could in principle be a normal red giant or 
supergiant reddened by over 70 magnitudes of visual extinction. This 
extinction is unlikely to arise from diffuse material in N159, in addition there is no obvious concentration in the CO maps coincident with the line 
of sight toward star No.~134. Sources in the upper right could in principle 
be reddened red supergiants, although their SEDs will argue otherwise. 

Several stars can be identified as OB stars based on optical work. Among 
the brighter at optical wavelengths are Sk-69~254 (B5~Iab), LMC~X-1 
\citep[O7~III,][] {cow95} and Sk-69~257 (O9~III). \citet{deh92} list a 
number of early type stars in the vicinity of N159A. They classified 
several of these stars based on dereddened absolute $V$ magnitudes and 
$UBVRI$ colors. The spectral energy distributions of several of these early 
type stars are shown in Fig.\ref{tjj-f6}. Star No.~16 (the B supergiant) 
probably has an extended ionized wind that creates excess flux in the red 
and near-IR. \citet{deh92} identified two stars very close to one another, 
having spectral types O5 and O7, that they suggested power most of N159A. 
We identify these stars as star No.~16.2 in Table~\ref{spsc}, a blend in 
our {\it Spitzer} images. 

\citet{deh92} typically found a reddening of 1-2~mag at $V$ in the regions 
of N159 they studied. Reddening at this level is evident in the SEDs in 
Fig.~\ref{tjj-f6} at wavelengths $\le 1$~\micron. The effects of this 
moderate reddening are minimal at the longer wavelengths of our 
observations, where the SEDs should become Rayleigh-Jeans. The SEDs of the 
early type stars in Fig.~\ref{tjj-f6} begin to approach a Rayleigh-Jeans 
continuum at the longer wavelengths, but some still show a small residual infrared 
excess. This is probably due to a combination of free-free emission from a 
wind and adjacent unresolved nebulosity contaminating our photometry. 

\subsection{Red Giants \label{rgs}}

The SEDs of a selection of luminous cool stars is shown in 
Fig.~\ref{tjj-f7}. To the best of our knowledge, only stars No. 72 (LMC V3938) and No. 147 (KDMK 5550) are optically identified red giants, and are classified as a carbon stars. The other stars were selected based on their colors in the 
color-magnitude diagram (Fig.~\ref{tjj-f5}). All of the stars presented in 
Fig.~\ref{tjj-f7} are similar, with energy distributions peaking between 1 
and 2~\micron, typical of cool stars. 

The bolometric correction to our [3.5] photometry is $\simeq +3.0$ for an 
M5 star and $\simeq +2$ for a G0 star \citep[see][]{bes98}.  We have used 
these bolometric corrections to construct the location of the AGB Limit 
(M$_{bol} = -7.1$) in the upper left corner of Fig.~\ref{tjj-f5}. The AGB 
limit represents the maximum luminosity possible for red giants with 
degenerate cores. The main sequence progenitors for these stars must be 
less than about 8-10~M$_{\odot}$. Any late type stars above this limit must 
be true supergiants with main sequence progenitors greater than 
8-10~M$_{\odot}$. We find no unreddened stars above the AGB limit. In fact, 
the brightest late type stars (No.~3 and 91, Table~\ref{spsc}) are over 
1~mag below the AGB limit. This is in contrast to 30 Doradus, where 
\citet{MH81} found numerous cool stars above and just below the AGB limit, 
which they identified as true red supergiants with very massive main 
sequence progenitors. 

All of these stars could be field red giants in the LMC or the occasional foreground halo dwarf. If the brightest of the red stars Nos.~3 and 91 were true supergiants, they would have minimum main 
sequence progenitors $10-12$~M$_{\odot}$ with ages of $10-20$~Myr. If, as we 
suspect, they are field red giants, they come from main sequence 
progenitors with a lifetime exceeding 30 million years and could easily be 
over $10^{8}$ years old.  \citet{deh92} discuss the stars in their optical 
sample with red intrinsic colors and conclude they are field giants with 
some contamination from halo dwarfs.  \citet{mey04} explore the possibility 
that their brighter 2~$\micron$ sources are late type stars in the LMC, but 
also find no stars above the AGB limit.  

Because IRAC mosaic images contain an off-source extension, we can compute 
the star counts in our band~1 images on and off N159. Star counts were 
conducted by dividing the band~1 mosaic into two separate rectangular 
sections, one taking in all of N159 ($44~\sq^{\prime}$) and the other section 7.2\arcmin \ south 
that was 33\% larger ($59~\sq^{\prime}$). The two 
protostars and 4 point sources that have the colors of HII regions were 
removed from the N159 (north) sample. The two brightest objects in the 
other (south) sample were removed due to the presence of extended 
nebulosity. A comparison of the star counts down to $[3.5] = 15$~mag. is 
shown in the histograms in Fig.~\ref{tjj-f8}. The vertical axes have been 
scaled to compensate for the difference in surface area on the sky between 
the two samples. It is clear from Fig.~\ref{tjj-f8} that there is no 
statistically significant excess population of stars at 3.5~$\micron$ in or near N159. The 
only possible difference between the two fields is the trend in the 
southern sample to have more stars with [3.5]~$\approx 10$~mag. 

The nature of the red stars in N159 is important for an understanding of 
the star formation history of N159. \citet{MH81} found a population of true 
supergiants in the 30~Doradus region that were distributed differently than 
the luminous blue stars. They showed that this indicated a previous episode 
of star formation about $5 \times 10^{7}$ yrs-ago. However, N159 is not as rich in massive stars as 30 Doradus and the red supergiant phase is short-lived. Consequently, we need to ask how many supergiants we would expect to see if most of the massive stars in N159 formed $\ga 10^{7}$ yrs ago. Stars betweem about 10 and 35 M$_{\odot}$ ($\sim$ B2 - O8) will evolve into true red supergiants and roughly one supergiant will be visible for every 10 OB stars present if the association is old enough for the some of the stars to have evolved off the main sequence to the red supergiant phase \citep{hum84}. \citet{deh92} tabulate 14 {\it optically visible} OB stars in the N159A and N159AN region, an area which accounts for about 40\% of the total far-IR luminosity of N159 \citep{jon86}. Keeping in mind there are certainly many OB stars too heavily reddened for Deharveng and Caplan to see (all of N159AN for example), there must be well over 40 stars B2 and earlier in all of N159. If most of the present massive star population in N159 was $10^{7}$ yrs old, we would expect to see four or more true red supergiants. We do not find any in N159, indicating most of the star formation has taken place more recently. We will argue below that N159 is, in fact, no older than $1-2$ Myr, and is forming new stars at the present epoch. Of course, due to small number statistices, a minor burst of star formation at an earlier epoch can not be completely ruled out by the lack of red supergiants. 

\subsection{HII regions \label{hii}}

Some of the regions bright in the IRAC band 4 image have very compact, 
nearly unresolved sources and are associated with peaks in the radio map. These include the Papillon (No.~98), N159S 
(No.~78) and Nos. 8 and 32. Not listed in Table~\ref{spsc} are two 
other point-like sources that are too embedded in nebulosity to permit 
accurate photometry. These point-like sources are at RA = $5^{h} 39^{m} 
38\fs05$, Decl. = $-69\degr 45\arcmin 27\farcs0$ (J2000), the brightest 
source at 7.9~\micron \ in N159AN, and at RA = $5^{h} 39^{m} 37\fs61$, Decl. 
= $-69\degr 46\arcmin 12\farcs3$ (J2000), the western peak in N159A, both discussed in \S\ref{genmorph}. 

Although the HII regions in N159 are generally extended objects, we can 
treat them as single entities by placing a synthetic aperture over them on 
the IRAC images. The results of this exercise are given in 
Table~\ref{aper-table}, where the aperture size is given along with the 
computed fluxes in Jy. We measured the fluxes on the IRAC images as they 
were oriented in the focal plane at the time of the observations. 
Consequently, the orientation of the horizontal axis of the aperture is not 
East-West on the sky, but at an angle of 56.5\degr \ East of North. 

The integrated SEDs of three HII regions are plotted in 
Fig.~\ref{tjj-f9}. Note that all of these objects have nearly identical SEDs, 
with an initial drop from $3.5 - 4.5$~\micron , then a steep rise to 
7.9~\micron . The SEDs for the HII regions in N159 are very similar to the SEDs for dusty galactic HII regions and other LMC HII regions such as Henize 206 \citep{gor04}. There is nothing in our observations that indicates a difference 
between the SEDs of HII regions in N159 and in the Milky Way. 

\subsection{The Protostars \label{protos}}

\citet{gat81} used $JHKL^\prime$ photometry to discover a very red source 
in N159. This source had no evidence of Br$\gamma$ emission or CO 
absorption in its 2~\micron \ spectrum, indicating it was not a hidden HII 
region nor a dusty oxygen rich red giant. \citet{gat81} proposed that this 
source was a protostar (later designated as P1). Subsequently \citet{jon86} 
found a second candidate, P2 near N159A. Both of these objects have 
received attention in later work. 

\citet{mey04} observed P1 (designated No.~210 in their tabulation) in their 
near-IR survey of the region surrounding the Papillon They discuss the 
possibility this star is a highly reddened O star, but find the necessary 
luminosity excessive. \citet{deh92} discuss P2 in the context of their 
optical CCD photometry. They suggest P2 is associated with star No.~217 in 
their list, which they classify as B1~V based on a computed M$_{V}$. We 
will show below that this association is probably incorrect. 

\citet{com98} present ISOCAM 15~$\micron$ images of N159A and find two 
sources, designated 1501E and 1501W. If we adjust their coordinates 
5$\arcsec$ north, these two sources nicely match up with P2 (1501E) and the 
western peak in N159A discussed in \S\ref{hii} and \S\ref{genmorph} (1501W). The western source, 
1501W, is brighter than P2 in our band~4 image and in the ISOCAM 
15~$\micron$ image. \citet{com98} find no peak in radio emission at the 
location of P2 at 1.4 and 2.4~GHz, but proceed to associate all of the 
far-IR luminosity for N159A with P2 and classify it as an ultracompact HII 
region. We believe this is incompatible with the SED of P2 and the location 
of the peak intensity in the far-IR and radio emission maps. 

Let us consider the nature of P2 by examining the
the 5~GHz contours \citep{hun94} in the vicinity of N159A as 
shown in Fig.~\ref{tjj-f3}. There is no radio peak at P2, but rather a peak 
to the southwest of P2, and another peak at the location of source 
1501W to the west of P2. Careful examination of the far-IR map \citep{jon86} shows that the 
100~\micron \  emission peaks southwest and west of P2 as well, in close 
correspondence with the radio. The radio peak west of P2 is clearly 
associated with source 1501W, which is also coincident with star No.~230 in 
\citet{deh92b}. This is certainly a compact HII region with one or more OB stars, some of which may be optically visible. The radio peak to the 
southwest of P2; however, is most likely associated with ionizing radiation from the O5 and O7 
stars identified by \citet{deh92}, {\it not} P2. 

The SEDs for P1, P2 and No. 9 are shown in Fig.~\ref{tjj-f10}. The near-IR data is 
from 2MASS, which is consistent with the {\it JHK} photometry of 
\citet{mey04} and \citet{jon86}. The warm component of the SEDs of P1 and 
P2 can be fit with a simple 880~K blackbody. We find that only No. 9, No. 134 and possibly No. 2.1 have an SED similar to the two protostars. Without spatially 
resolved observations at longer wavelengths, we can not determine if there 
is significant emission longward of 20~\micron. For P2, however, we can use 
the ISOCAM observations at 15~\micron \ from \citet{com98}, assuming the 
shift in coordinates mentioned above. Although we could not extract point 
source photometry for 1501W from our IRAC images due to excessive 
surrounding nebulosity, examination of the source in a square 
11\arcsec~$\times$~11\arcsec \ synthetic aperture reveals that 
it has the SED of 
an HII region (Fig.~\ref{tjj-f9}), not the SED of P1 or P2.  

The [3.5]-[4.5] and [5.7]-[7.8] colors of the sources in the N159 
field sort out into different regions in the color-color diagram shown 
in Fig.~\ref{tjj-f11}. The OB stars are far too faint in band~4 to 
be measured, and are not represented in Fig.~\ref{tjj-f11}
The red giants, which we have argued are not part of N159, are well 
separated from the other sources in this color-color plane. 
\citet{lallen04} present recent {\it Spitzer} observations 
of Class~I and Class~II protostars in several Galactic star forming 
regions. They plot the IR colors of these sources along with 
theoretical models of {\it low mass} YSOs discussed in \citet{dal03}. The regions 
in the {\it Spitzer} color-color diagram occupied by their observed 
Class~I and Class~II YSOs are shown in Fig.~\ref{tjj-f11}. 
The reddening vector is taken from \citet{lallen04} for the case of a 
flat spectrum source. We can see in Fig.~\ref{tjj-f11} that P2, No.~9 and 
No.~134 have the colors of low mass YSOs. The HII regions (some from Table~\ref{spsc}
and some from Table~\ref{aper-table}) plot in a different part of the 
diagram, although there could certainly be some overlap. Also, a compact HII region reddened by A$_{V} \ga 20$ will have colors similar to Class I YSOs. The lack of clear association of radio and far-IR emission with P2 (or No.~134) suggest they are high mass versions of true Class I YSOs, not reddened HII regions. We can not make the same claim for No. 9, which is embedded in radio and far-IR luminosity. 

Note that P1, the original protostar, has [5.7]-[7.9] colors far bluer than 
expected, based on our suggestion it is a Class~I YSO, or even a Class~II YSO. Source 2.1 has [5.7]-[7.9] colors intermediate between P1 and P2. Sources 2.1 and 3.1 are very close to one another within a patch of nebulosity southwest of the main N159 complex (designated {\it maser} in Fig.~\ref{tjj-f1}). No. 3.1 corresponds to the position of the H$_2$O maser given in \citet{hun94}, but has a rapidly dropping SED to longer wavelenghts, and could not be measured at 5.7 and 7.9~$\micron$. No. 2.1 corresponds to an MSX point source and has colors suggestive of a Class~II YSO, but would have to be more luminous (L$\approx 300 L_{\odot}$) than typical Galactic T Tauri stars. The nebulosity surrounding these two sources dominates the fluxes listed in Table~\ref{aper-table}, which clearly show that the maser region has the colors of an HII region. Perhaps No. 2.1 is an embedded Herbig Be star, still with its circumstellar disk, and creating a small HII region.

In Fig.~\ref{tjj-f12} we show the location of stars optically identified by 
\citet{deh92b} overlying a contour map of our band~1 image. Several stars they
found have corresponding point sources at 3.5~\micron. 
These are either O stars or stars not typed by \citet{deh92}. The O5 and O7 
stars are clearly present as a blend in our IRAC image. These two stars 
probably account for a significant fraction of the ionizing radiation in 
N159A, and are probably the source of the radio and far-IR radiation that 
peaks southwest of P2 \citep{deh92}. None of the stars classified as B stars by 
\citet{deh92} can be distinguished from the nebulosity in our band~1 image, 
presumably because they are below our detection threshold in this confused 
region. Note that the B1~V star near P2 is clearly not coincident in 
position, and is probably not associated with P2. We conclude that P2 is a Class~I YSO that may produce some ionizing radiation, but is not an embedded O star.

P1 is in a more isolated region than P2, and also shows no clear evidence for 
an optical component or any associated radio or H$\alpha$ emission. The SEDs of P1 and P2 can be fit by models of Class~I YSOs 
\citep{ada87, whi03}, although the flux from P1 is weaker at 
7.9~\micron \ than most of the models presented in these papers. This is 
evident by the location of P1 in the color-color diagram (Fig.~\ref{tjj-f11}), 
much bluer in [5.7]-[7.9] color than expected. We do note, 
however, that Elias 29 in Taurus \citep{eli78}, a low mass Class~I YSO, 
does drop in flux from 5 to 10~$\micron$, similar to P1.

Since no far-IR peak is coincident with P1 or P2, the two protostars can not be 
contributing a major fraction of the luminosity of N159. We do not believe 
they are ultracompact HII regions or some other form of heavily embedded O 
stars with luminosities well above $10^{4}$~L$_{\odot}$. By analogy with the Class~I YSO KW in M17 \citep{KW73, nie01}, we assign 
a luminosity of a few $\times10^{3}$~L$_\odot$ for P1 and P2. These stars 
will evolve into a B2 or B3 main sequence star. Whatever the details of 
the nature of these two stars, they are very young, $\approx 10^{5}$~yrs-old. 
Infrared spectroscopy will be necessary to more accurately classify all 
of the possible protostars.

The peculiar colors of P1 force us to mention one caveat with regards 
to its classification. There is still the possibility that this star is 
a very cool carbon star or some other object that has an SED well fit 
by a $\approx 900$~K blackbody. This would require effective temperatures 
significantly cooler than the identified carbon stars in the N159 
field (Nos.~72 and 147). It is also possible P1, along with No. 2.1 are higher luminosity versions of Class~II YSOs. We are confident, however, that P2 is a Class~I protostar.

\section{DISCUSSION \label{disc}}

The SED of the entire N159 complex is shown in Fig.~\ref{tjj-f13} along 
with the SED of W51 \citep{sie91} scaled to correspond to the distance of 
the LMC. The far-IR data for N159 is taken from \citet{jon86}. The 
10-30~\micron \ data is taken from the {\it Infrared Satellite
Astronomical Observatory} (IRAS) and MSX observations. The total 
luminosity of N159 can be computed by integrating under the SED over all IR 
wavelengths. This assumes that there is no significant leakage of 
non-ionizing UV and optical radiation. \citet{jon86} investigated this 
possibility by comparing the observed UV luminosity of N159 to the far-IR 
luminosity and concluded that N159 was effectively thermalizing almost all 
of its stellar power. Missing from their analysis, however, were 
observations from $3-50~\micron$ which clearly show that there is 
significant dust at temperatures warmer than the single dust temperature 
derived by \citet{jon86} from the 50 and 100~\micron \ KAO observations. 

We find the integrated luminosity of the entire N159 complex to be $\approx 9 
\times 10^{6}$~L$_{\odot}$, roughly equivalent to 11 O5 stars \citep{pan73}. 
Using standard formulas and the radio luminosity of N159 at 6~cm 
\citep[2~Jy:][]{mcgee72, jon86}, the ionizing 
photon rate necessary to create the 
observed free-free emission is $Log(N_{\gamma}\ s^{-1}) = 50.6$. Note that 
some authors have incorrectly used the integrated radio flux at 408~MHz 
tabulated in \citet{cla76} which includes a much larger area than N159, 
including all of N160 to the north. Our result is compared to other HII 
regions, both in the LMC and in the Milky Way, in Fig.~\ref{tjj-f14}. The 
solid line in Fig.~\ref{tjj-f14} represents a hypothetical cluster with a 
normal IMF ending with the spectral type at the fiducial point on the line. 
The basic parameters for this relationship were take from \citet{ick80}, 
which should be characteristic of dusty Galactic HII regions \citep{lig83}. Note that 
dusty HII regions W51 and M17 fall close to this relationship. 

HII regions located well to the left of this line are either leaking 
significant amounts of luminosity at near UV and optical wavelengths or are 
producing more ionizing radiation than expected for their luminosity. N59A for example, is probably leaking radiation that is not being 
thermalized by dust absorption \citep{jon86}. Since N159 is converting most 
of its stellar radiation into IR flux, the IR luminosity should represent 
the total luminosity of the region. \citet{jon86} derived a lower 
luminosity for N159 than we find in this work and used different stellar 
parameters to build the theoretical relationship for a normal IMF cluster. 
Their analysis found N159 under-luminous for its derived ionizing flux, 
prompting them to hypothesize a top-heavy IMF for the LMC. As can be seen 
from the location of N159 in Fig.~\ref{tjj-f14}, the luminosity and 
ionizing flux for N159 is entirely consistent with the expected trend for 
HII regions in the Milky Way, and there is no need to invoke a top-heavy 
IMF. 

Unlike most galactic HII regions, which tend to have dense luminous cores, 
N159 is composed of about four HII regions separated from 
one another by 10-40~pc, 
each of which is roughly equivalent to in luminosity to $0.75 \times$ a 
typical M17 HII region. These are the Papillon, N159A, N159AN, and the 
central radio peak discussed in \S\ref{hii}. Using our new {\it Spitzer} 
observations, the 843~MHz radio map \citep{mil84}, and the far-IR map 
\citep{jon86}, we find the luminosity and ionizing flux from N159 is 
consistent with three or so O5/O6 stars located at the Papillon, N159A, and 
N159AN (each), and the equivalent of three or so O6/O7 stars at the center of 
the nebula plus contributions from less massive OB stars throughout the 
region. N159S makes only a minor contribution. Using the Z = 0.008 
evolutionary tracks in \citet{mey94} and the tabulation of O star 
properties by \citet{pan73}, the HII regions making up the N159 complex are 
less than 2~Myr-old. 

Are these separate HII regions that make up the N159 complex closely 
related, perhaps triggered by a common event? Or, are they relatively 
separate star formation events located within a 50~pc diameter volume 
simply because there is a significant amount of molecular gas 
nearby? A link between star formation in the LMC and the formation of large bubbles has been discussed by numerous authors. \citet{efr98} argue for a star formation history in LMC4/Constellation III triggered by the dynamic effects of superbubbles produced by previous generations of massive stars. \citet{ole01} use an extensive data set to study one area within LMC4 and conclude that LMC4 is made up of overlapping bubbles that have initiated past star formation events and that this process will continue into the future. We will argue that we are seeing the earliest stage in this process in N159, with the formation of its first OB driven bubble.

The local peak in radio emission at the center of the nebula 
is more diffuse and lower in surface brightness than the Papillon, 
N159A, or N159AN (\S\ref{genmorph}). The higher 
resolution 5~GHz map shows this emission breaking up into a 
long arc extending around the north, east and south of the 843~MHz peak. At 
this location in our IRAC band~1 image is a blend of two or more sources 
with a combined [3.5] = 12.86~mag (No.~53.1 in Table~\ref{spsc}). The 2MASS 
point source catalog lists 1 source at this location. The higher resolution 
images of \citet{mey04} find two sources within 1\arcsec \ of our blend, 
No.~7 and No.~15 in their tabulation. By analogy to the O5+O7 blend in 
N159A ([3.5] = 12.36~mag, star No.~16.2 in Table~\ref{spsc}, discussed in 
\S\ref{cmd}) these stars could be O stars $\approx 0.5$~mag fainter, roughly 
O6+O8, and the primary source of ionizing flux for the central radio peak. 
However without spectra this assertion can not be confirmed. Nevertheless, the radio observations show that ionizing flux indicative of several middle O stars is being produced at this 
location. 

These results suggest that several O stars were born at the center of N159 
about $1-2$~Myr-ago, and have blown a bubble in the ISM. Fig.~\ref{tjj-f15} 
shows a gray scale representation of our IRAC band~4 (7.9~\micron) image 
over which a circle about 40~pc in diameter, centered on the sources at the 
central radio peak, is drawn to help illustrate our proposed scenario. The 
distribution of molecular gas discussed in \S\ref{genmorph} 
(Fig.~\ref{tjj-f4}) peaks at the periphery of this bubble. The high surface 
brightness (young) HII regions are located around the periphery as well. 
Studies of large spherical voids in the interstellar
medium (ISM) using 21~cm observations of 
neutral hydrogen suggest ages of $\approx 1-2$~Myr for a bubble the size and 
location we find for N159 \citep{oey96, mcc01}. Specifically, computations by \citet{oey96} show that DEM 301, which contains an OB population similar to the one we propose at the center of N159, would have blown a 40pc diameter bubble in about 1.5~Myrs. 

The morphology of N159 is 
consistent with an OB wind blown bubble, about $1-2$~Myr-old that has 
partially evacuated the central region of a molecular cloud and compressed 
the molecular gas at the periphery. This has initiated a new, very recent 
burst of star formation at several locations along the periphery of the 
bubble. The absence of any red supergiant population (see \S\ref{rgs}) implies there was no star formation episode involving tens of OB stars previous to the one that 
created the bubble. The presence of several relatively luminous Class~I YSOs indicates star 
formation is currently taking place in N159. 

The presence of LMC X-1 and the optically bright blue giants and 
supergiants may complicate this simple scenario. Using {\it Chandra} 
imaging, \citet{wil00} find faint soft X-Rays in the northeast sector of 
N159 which they associate with SNR 0540-6994, estimated to have taken place about $1-2 \times 10^4$ yrs ago. The location of the brighter portion of this X-Ray emission is indicated in Fig.~\ref{tjj-f15} and is not believed to be associated with LMC X-1. 
We find a ring of emission in our band 4 image about 19~pc in diameter that corresponds to the location of the soft X-Ray emission. The size of the supernova remnant is too small to have influenced star formation elsewhere in N159. 

Using the Geneva evolutionary tracks for Z=0.008 \citep{mey94} and published spectral types, we can estimate the main sequence progenitor mass for the optically bright blue stars in N159. LMC X-1 contains an O7~III giant \citep{cow95}, which, if unreddened, has an M$_{bol} \approx -8.8$. This corresponds to a main sequence progenitor of about 35M$_\odot$, or an age of less than $2\times 10^6$ yrs. Based on its B-V color, Sk-69 257 (O9~III) is reddened by about A$_V$ = 0.7 \citep{wal77}. This implies M$_{bol}$ = -9.9, corresponding to a main sequence progenitor of 55M$_\odot$ and an age of less than $1.4 \times 10^6$yrs.

Only the B5~Iab supergiant may be older than the 2~Myr age we associate 
with the wind blown bubble, yet still young enough to have been part of a 
relatively recent burst of star formation. Assuming spectral type B5, we 
find  E(B-V) = 0.36~mag, A$_{V} = 1.0$~mag, and therefore M$_{V} = -
7.6$~mag. This corresponds to M$_{bol} = -8.5$~mag. and the main sequence progenitor for this star 
would only be about 25~M$_{\odot}$, with an age no less than 2.5~Myr. With 
this one exception, none of these ages we find for the optically visible OB 
stars are inconsistent with the simple scenario we have proposed. 

\section{CONCLUSIONS \label{concl}}

Based on our new {\it Spitzer} observations of the protostars in N159, we 
classify P2 as a Class~I YSO with an age of $\approx 10^{5}$~yrs and a 
luminosity of $3 - 5~\times~10^{3}$~L$_{\odot}$. The original protostar, P1 
is also likely a Class~I YSO (or perhaps a Class~II YSO), but its [5.7]-[7.9] color is not as 
red as typical Milky Way Class~I YSOs. Two new sources have the colors of 
Class~I YSOs and deserve attention in future observations. 

The total luminosity of the entire N159 complex 
is $\approx 9~\times~10^{6}$~L$_{\odot}$. 
The ratio of total radio flux density to 
total luminosity is consistent with typical dusty HII regions observed in 
the Milky Way. There is no evidence for an excess of ionizing photons 
indicative of a top heavy IMF. There is no evidence of a previous episode 
of significant star formation that would have produced core burning 
supergiants that would still be visible today.

Star formation in N159 appears identical to star formation in the Milky Way, in particular an association with molecular gas and the presence of relatively high mass protostars. N159 is somewhat different, however, in its large physical size. The entire complex is over 50pc in diameter with at least four separate concentrations of ionizing flux and mid-IR luminosity. We argue that all of these separate HII regions are related to one another, the result of triggered star formation on the scale of tens of parsecs. 

\acknowledgements
This work is based on observations made with the {\it Spitzer Space 
Telescope} which is operated by the Jet Propulsion Laboratory, California 
Institute of Technology under NASA contract 1407. Support for the authors 
was provided by NASA through contracts 1256406 and 1215746 issued by JPL/Caltech.


\newpage

FOR HIGH QUALITY FIGURES VISIT http://www.astro.umn.edu/\~~tjj/ \\
CLICK ON ~~~~~ \textbf{N159 Paper Figures}

\figcaption[tjjones.fig1.eps]{Gray scale image of N159 in {\it Spitzer} \ 
IRAC band 1 (3.5~$\micron$). Several objects discussed in the 
text are identified. 
\label{tjj-f1}} 

\figcaption[tjjones.fig2.eps]{False color image of N159 using {\it Spitzer} \ 
IRAC band 1 (blue), band 2 (green) and band 4 (red). Red giants appear blue due 
to CO absorption in band 2. Hot dust and PAH emission appear green and red. 
\label{tjj-f2}} 

\figcaption[tjjones.fig3.eps]{{\it Left panel} (a): Gray scale image of 
N159 overlain with the 5~GHz radio contours from \citet{hun94}. {\it Right 
panel} (b): Same as (a), but with the H$\alpha$ contours from 
\citet{hun94}. 
\label{tjj-f3}} 

\figcaption[tjjones.fig4.eps]{Gray scale image of N159 overlain with the 
3~mm CO contours from \citet{joh98}.
\label{tjj-f4}} 

\figcaption[tjjones.fig5.eps]{Color-Magnitude diagram of point sources in 
N159 from band~1 and band~2 {\it Spitzer} \ IRAC images. The numbers refer to 
sources listed in Table~\ref{spsc}. The spectral types along the horizontal 
axis mark the calculated [3.5]-[4.5] color magnitude for unreddend late 
type stars.
\label{tjj-f5}}

\figcaption[tjjones.fig6.eps]{Spectral Energy Distributions (SEDs) for 
selected early type stars. Numbers refer to sources listed in 
Table~\ref{spsc}. The probable O5+O7 blend is No.~16.2, and LMC X-1 is No.~14. 
\label{tjj-f6}} 

\figcaption[tjjones.fig7.eps]{Spectral Energy Distribution (SEDs) for 
selected late type stars. Numbers refer to sources listed in 
Table~\ref{spsc}. Note the SEDs peak between 1 and 2~\micron \ and note the dip 
in the IRAC band~2 (4.5~\micron) filter due to CO absorption in the 
atmosphere of the star at 4.6~$\micron$. 
\label{tjj-f7}}

\figcaption[tjjones.fig8.eps]{Star counts in {\it Spitzer}\  IRAC band~1 in N159 
and a region of sky 7.2$\arcmin$ south. The region including N159 covered 
$44\sq^{\prime}$ on the sky. The comparison region to the south 
covered $59~\sq^{\prime}$. The vertical axes have been scaled to take 
into account the different areas. 
\label{tjj-f8}} 
 
\figcaption[tjjones.fig9.eps]{Spectral Energy Distribution (SEDs) for HII 
regions listed in Table~\ref{aper-table}. The Papillon and Center regions 
have almost identical SEDs to N159AN and were not plotted. The area 
containing the maser source has an SED almost identical to 
N159S and is not plotted. 
\label{tjj-f9}}

\figcaption[tjjones.fig10.eps]{Spectral Energy Distributions (SEDs) for the 
two protostars and source No. 9. The dotted line shows a simple 
800~K black body spectrum. 
The data point for P2 at 15~\micron \ is taken from \citet{com98}. 
\label{tjj-f10}} 

\figcaption[tjjones.fig11.eps]{Color-Color diagram using the 
four IRAC bands. The locations of Class~I and II YSOs are taken 
from \citet{lallen04}. All point sources from Table~\ref{master} and some extended sources in Table \ref{aper-table} are plotted ({\it filled circles}) with some labeled. 
\label{tjj-f11}}

\figcaption[tjjones.fig12.eps]{Location of stars optically identified by 
\citet{deh92} overlaying a contour map our {\it Spitzer} \  IRAC band~1 image of 
N159A. {\it Filled circles} indicate tentative association of 
optical stars with 3.5~\micron \ sources. The identification numbers are 
taken from \citet{deh92}, not our numbering system, but can be cross 
referenced using Table~\ref{spsc}. Note that the protostar, P2 is not 
coincident with any optical star. 
\label{tjj-f12}}

\figcaption[tjjones.fig13.eps]{Spectral Energy Distributions (SEDs) of the 
entire N159 complex compared to the galactic HII region W51. The fluxes for 
W51 have been scaled to the distance of the LMC. 
\label{tjj-f13}}

\figcaption[tjjones.fig14.eps]{Plot of the ionizing photon rate, 
N$_{\gamma}(s^{-1})$, determined from the radio emission against the total 
luminosity derived from 1-100~\micron \ fluxes for a selection of HII 
regions. Note that N59A is also in the LMC. The line represents the computed 
relationship for a hypothetical cluster with a normal IMF ending in a 
single O star with the spectral type indicated on the line. This cluster is 
purely hypothetical as it contains fractions of a star at some spectral 
types. 
\label{tjj-f14}} 

\figcaption[tjjones.fig15.eps]{Gray scale image of N159 in the {\it Spitzer} \ 
IRAC band 1. A circle about 40~pc in diameter, centered on the sources near the 
central radio peak, illustrates the proposed wind blown bubble we 
hypothesize has initiated star formation at its periphery. Note how many 
young objects lie at the outer edge of this bubble. Also compare with the CO 
distribution shown in Fig.~\ref{tjj-f4}. 
\label{tjj-f15}}



\begin{deluxetable}{ccccccccccc}
\tablenum{1}
\tablewidth{0pt}
\tablecaption{IRAC Point Source Fluxes in $\mu$Jy \label{master}}
\tablehead{

\colhead{No.} & \multicolumn{2}{c}{RA (J2000) Decl.} & \colhead{F$_1$} & \colhead{$\epsilon_1$}
& \colhead{F$_2$} & \colhead{$\epsilon_2$} & \colhead{F$_3$} & \colhead{$\epsilon_3$} 
& \colhead{F$_4$} & \colhead{$\epsilon_4$}}
\startdata

0&84.833283&-69.764582&29600&85.1&15200&53.4&11800&76.6&11000&63.6 \\
1&84.847602&-69.747225&353&11.5&166&7.6&&&& \\
2&84.865127&-69.757576&160&9.1&147&6.6&&&& \\
2.1&84.871976&-69.788730&5050&38.7&6520&36.6&13500&61.20&11000&63.6 \\
3&84.872272&-69.770773&39600&57.0&20700&35.7&17400&56.4&17400&48.5 \\
3.1&84.875649&-69.789187&6650&44.0&7800&39.9&&&& \\
4&84.885455&-69.773384&1070&12.8&684&9.0&&&& \\
5&84.887750&-69.766636&704&11.8&494&8.5&&&& \\
6&84.893753&-69.759816&575&12.7&599&9.9&&&& \\
7&84.895446&-69.779635&786&11.4&471&8.1&&&& \\
8&84.900225&-69.767821&7800&26.5&8300&21.9&30500&68.6&90700&80 \\
9&84.904689&-69.760189&9520&31.6&21500&33.7&61700&90.2&132000&90.7 \\
10&84.905032&-69.749125&4190&21.2&2070&13.3&&&& \\
11&84.906539&-69.745679&558&10.7&430&8.2&&&& \\
12&84.907488&-69.761284&5470&22.1&3820&16.1&&&& \\
13&84.907862&-69.739862&2280&16.6&1120&10.6&&&& \\
14&84.912176&-69.743215&1720&15.0&1150&11.3&&&& \\
15&84.915065&-69.789524&1140&15.5&628&10.4&&&& \\
16&84.917086&-69.742745&11800&28.2&8810&21.4&5950&31.3&3220&60.00 \\
16.1&84.917112&-69.731142&4100&21.2&2639&200.0&&&& \\
16.2&84.917527&-69.771995&3160&18.0&&&&&& \\
17&84.920176&-69.775445&2600&16.4&3880&16.2&9880&44.2&28400&49.8 \\
18&84.921482&-69.736015&666&9.4&397&7.3&&&& \\
19&84.922616&-69.764758&1500&12.6&904&8.8&&&& \\
20&84.923365&-69.734688&485&9.1&295&6.5&&&& \\
21&84.924819&-69.769968&48800&55.1&84600&61.7&180000&135.0&302000&121 \\
22&84.924974&-69.798697&17700&47.4&10400&31.9&9510&49.1&6160&36.2 \\
23&84.926247&-69.783607&1650&13.3&799&8.8&&&& \\
24&84.926557&-69.757771&7800&23.7&4020&15.2&&&& \\
25&84.926730&-69.753047&1850&16.0&922&10.7&&&& \\
26&84.927280&-69.782304&2170&14.1&1080&9.2&&&& \\
27&84.932593&-69.761150&744&10.4&1220&9.9&&&& \\
28&84.933262&-69.749352&1730&13.6&1120&10.1&&&& \\
29&84.934104&-69.787148&499&10.3&270&7.3&&&& \\
30&84.935497&-69.743317&576&10.8&502&8.2&4310&32.0&& \\
31&84.936093&-69.737354&835&10.0&474&7.2&&&& \\
32&84.938616&-69.747381&2200&14.6&3300&14.1&11200&43.8&37700&56.4 \\
33&84.939370&-69.800695&5550&23.2&3900&16.8&3770&30.3&3130&29.9 \\
33.1&84.940823&-69.791981&978&11.9&324&7.9&&&& \\
34&84.941584&-69.753896&1030&11.0&575&8.0&&&& \\
33.2&84.942422&-69.790947&408&9.4&115&7.4&&&& \\
35&84.943332&-69.758697&810&9.6&509&7.4&&&& \\
36&84.945481&-69.775241&801&10.2&457&7.0&&&& \\
37&84.946334&-69.746717&1850&14.2&1070&9.9&&&& \\
38&84.948029&-69.732866&940&11.6&704&8.2&&&& \\
39&84.951922&-69.798016&469&9.9&248&6.7&&&& \\
40&84.952080&-69.726893&692&9.3&403&6.9&&&& \\
41&84.954453&-69.795089&1200&12.8&751&8.7&&&& \\
42&84.955312&-69.798620&377&9.2&209&10.1&&&& \\
43&84.956728&-69.756977&2160&13.2&1520&10.5&&&& \\
44&84.958251&-69.815450&1110&14.1&619&13.9&&&& \\
45&84.958392&-69.765183&1240&10.6&688&7.7&&&& \\
46&84.961990&-69.800116&241&10.8&176&6.5&&&& \\
47&84.965898&-69.809042&116&9.0&97&5.0&&&& \\
48&84.966078&-69.775721&25900&40.0&13400&25.4&9450&43.1&5430&28.9 \\
49&84.967299&-69.799960&523&9.4&273&6.5&&&& \\
50&84.967789&-69.747070&670&10.7&402&7.4&&&& \\
51&84.968256&-69.791491&175&8.3&123&5.9&&&& \\
52&84.969227&-69.810467&975&9.7&597&9.7&444&15.9&& \\
53&84.969730&-69.763073&1100&10.3&676&7.8&&&& \\
53.1&84.970800&-69.754900&2000&200.0& & &&&& \\
54&84.972574&-69.722934&393&8.7&603&7.6&&&& \\
55&84.973658&-69.759921&899&10.4&578&7.9&&&& \\
56&84.973754&-69.741518&1300&12.4&733&8.9 \\
57&84.976287&-69.768741&1290&10.8&747&7.9 \\
58&84.976682&-69.800360&105&7.2&98&5.5 \\
59&84.978377&-69.725337&849&10.9&501&7.7 \\
60&84.981385&-69.801131&108&6.7&91&5.1 \\
61&84.981557&-69.737821&804&10.4&500&8.0 \\
62&84.982210&-69.807182&130&6.3&104&6.0 \\
63&84.983726&-69.771865&2340&13.0&1260&9.2 \\
64&84.986388&-69.805764&181&7.0&182&8.4 \\
65&84.991237&-69.729958&891&10.4&550&7.7 \\
66&84.992957&-69.724098&1740&13.2&985&9.3 \\
67&84.993779&-69.788604&291&7.4&173&5.5 \\
68&84.994608&-69.802427&460&8.1&295&6.5 \\
69&84.994773&-69.744041&2190&13.0&1300&9.4 \\
70&84.995066&-69.734479&2400&14.9&1500&10.8 \\
71&84.995082&-69.769322&683&8.8&298&6.4 \\
72&84.995299&-69.804636&16900&32.7&12900&35.0&10100&41.0&6250&35.9 \\
73&84.997660&-69.757310&48900&50.0&59700&60.0&91400&100.0&57000&51.8 \\
74&84.999726&-69.748790&1200&11.0&802&8.3&&&& \\
75&85.000743&-69.803451&561&8.5&303&11.0&&&& \\
76&85.001368&-69.770022&861&9.2&595&7.8&&&& \\
77&85.002347&-69.768257&881&9.8&517&7.2&&&& \\
78&85.003217&-69.787037&18700&33.9&22900&32.3&41100&64.9&99300&70.2 \\
79&85.003698&-69.737406&3050&16.7&1780&11.7&&&& \\
80&85.003786&-69.799625&3150&17.0&1570&11.8&&&& \\
81&85.004621&-69.767397&717&9.3&449&7.1&&&& \\
82&85.006276&-69.716778&2710&14.5&1600&10.0&1050&100.0&& \\
83&85.007177&-69.743928&677&9.9&422&7.8&&&& \\
84&85.007568&-69.729639&7020&22.5&3820&14.8&&&& \\
85&85.007959&-69.782087&4000&17.2&1910&11.9&1670&19.0&& \\
86&85.008754&-69.809427&661&12.5&341&11.8&&&& \\
87&85.010208&-69.762824&867&8.9&603&7.0&&&& \\
88&85.010648&-69.776329&1390&11.3&675&7.7&&&& \\
89&85.011102&-69.792783&377&8.0&214&6.5&&&& \\
90&85.011600&-69.749901&691&10.6&1230&10.2&&&& \\
91&85.013634&-69.712708&37400&54.6&19000&29.6&11700&38.4&10100&29.7 \\
92&85.015390&-69.805283&1390&15.7&755&15.0&&&& \\
93&85.015390&-69.786127&527&8.4&330&6.2&&&& \\
94&85.015802&-69.722600&2890&15.2&1570&10.3&&&& \\
95&85.017325&-69.731105&539&9.4&343&7.0&&&& \\
96&85.017479&-69.749666&1480&12.6&1030&9.8&&&& \\
97&85.018165&-69.776570&575&8.4&268&8.4&&&& \\
98&85.018416&-69.743778&19500&32.4&29700&33.2&92700&90.0&184000&115 \\
99&85.022372&-69.761990&632&8.1&473&6.5&&&& \\
100&85.022756&-69.803966&550&9.3&311&7.9&&&& \\
101&85.023838&-69.750616&3440&16.1&2190&12.4&&&& \\
102&85.025467&-69.731534&823&10.6&468&7.7&&&& \\
103&85.027021&-69.708022&1880&17.8&1150&12.3&&&& \\
104&85.028459&-69.739669&1350&11.7&670&8.5&&&& \\
105&85.028791&-69.761985&813&9.5&580&7.2&&&& \\
106&85.030017&-69.718599&22300&42.8&14400&30.1&12100&45.4&9890&35.5 \\
107&85.030092&-69.723631&5150&19.3&3350&13.7&2190&23.9&& \\
108&85.030871&-69.778227&1600&12.1&812&7.9&&&& \\
109&85.031230&-69.760312&2380&13.4&1320&9.5&&&& \\
110&85.032559&-69.775597&1000&10.0&583&7.3&&&& \\
111&85.035517&-69.780830&2060&13.2&995&8.5&&&& \\
112&85.037321&-69.712831&2530&16.3&1470&11.3&&&& \\
113&85.038097&-69.790296&940&9.5&548&6.6&&&& \\
114&85.040773&-69.777190&2270&13.5&1150&9.1&1010&20.1&& \\
115&85.045699&-69.777491&499&8.4&303&6.0&&&& \\
116&85.048040&-69.793138&1950&11.1&1140&9.8&827&14.8&& \\
117&85.050629&-69.785625&660&8.6&435&6.2&&&& \\
118&85.054606&-69.775734&2380&13.5&1380&9.3&1010&18.6&& \\
119&85.055884&-69.767687&491&7.8&300&6.4&&&& \\
120&85.056346&-69.758106&364&6.7&171&6.4&&&& \\
121&85.056500&-69.730653&8940&28.3&4530&15.5&3610&26.3&& \\
122&85.057502&-69.746397&1190&10.9&766&8.2&&&& \\
123&85.058323&-69.796863&766&10.3&375&8.5&&&& \\
124&85.058504&-69.787648&508&8.1&266&6.9&&&& \\
125&85.063696&-69.768976&4750&18.5&2500&12.0&1810&20.1&898&17 \\
126&85.065194&-69.779336&387&7.8&252&7.0&&&& \\
127&85.065342&-69.774513&120&5.5&74&4.2&&&& \\
128&85.066689&-69.730716&804&11.0&510&8.0&&&& \\
129&85.068327&-69.737033&479&8.5&270&6.8&&&& \\
130&85.070641&-69.781416&372&7.2&197&6.5&&&& \\
131&85.071127&-69.769250&797&9.1&490&6.6&&&& \\
132&85.072542&-69.773314&368&7.1&238&5.9&&&& \\
133&85.073338&-69.782863&1280&10.8&778&7.4&481&10.6&& \\
134&85.079663&-69.746004&490&9.5&1100&10.0&3070&30.5&4800&34.3 \\
135&85.079763&-69.790679&344&7.1&236&7.2&&&& \\
136&85.080613&-69.766480&322&7.5&280&6.2&&&& \\
137&85.081615&-69.769507&429&7.5&260&6.0&&&& \\
138&85.085311&-69.790994&296&7.3&236&49.7&&&& \\
139&85.086167&-69.757984&1260&10.6&740&7.5&&&& \\
140&85.092653&-69.780041&138&6.1&90&4.2&&&& \\
141&85.093947&-69.757859&734&8.6&461&6.4&&&& \\
142&85.096047&-69.744059&1080&12.6&629&8.7&&&& \\
143&85.099445&-69.787923&2250&15.3&1340&13.0&812&18.0&382&18.4 \\
144&85.100291&-69.781389&681&7.7&377&6.1&&&& \\
145&85.101043&-69.770367&132&6.8&91&4.2&&&& \\
146&85.102115&-69.758051&1250&10.5&739&7.3&518&12.6&& \\
147&85.105899&-69.779225&19300&31.5&11000&22.8&9280&28.8&8180&24.8 \\
148&85.107911&-69.770784&385&7.1&170&5.8&&&& \\
149&85.110000&-69.747023&430&9.3&250&7.7&&&& \\
150&85.112048&-69.759260&388&8.0&192&7.3&&&& \\
151&85.119080&-69.759522&5300&22.2&2870&14.4&2370&22.3&1710&18.6 \\
152&85.122867&-69.763727&1780&12.5&1010&9.4&802&16.9&& \\
153&85.125174&-69.776790&2490&12.3&1460&9.5&1120&17.1&574&13.6 \\
154&85.127148&-69.750906&145&9.8&78&6.2&&&& \\
155&85.131513&-69.774730&668&7.5&446&6.1&331&10.5&& \\
156&85.131844&-69.770181&1110&10.1&635&7.3&501&13.2&& \\
157&85.134568&-69.773123&145&7.2&122&5.0&&&& \\
158&85.137341&-69.769945&653&9.7&406&7.0&279&11.6&& \\
159&85.140820&-69.775138&971&9.4&591&9.5&442&15.3&& \\
160&85.143520&-69.770623&1340&12.4&752&8.5&543&13.7&& \\
161&85.147652&-69.761446&854&12.9&523&9.2&&&& \\
\enddata
\end{deluxetable}

\begin{deluxetable}{ccccccccccccccc}
\tabletypesize{\scriptsize}
\rotate
\tablenum{2}
\tablewidth{0pt}
\tablecaption{Selected Sources in N159 \label{spsc}}
\tablehead{
\colhead{No.} & \colhead{2MASS ID} & \colhead{Other ID} & \colhead{B} & 
\colhead{V} & \colhead{R} & \colhead{I} & \colhead{J} & \colhead{H} &  
\colhead{K$_{s}$} & \colhead{[3.5]} & \colhead{[4.5]}&  \colhead{[5.7]} & 
\colhead{[7.9]} & \colhead{Notes}}
\startdata

0&05391991-6945525&&&&17&15.2&12.82&11.30&10.59&9.93&10.18&9.99&9.40& \\
2.1&05392957-6947212&DCL46, N159H, MSXG280.2580-31.5245&19.83&19.14&18.83&18.45&15.3
7&13.58&12.65&12.27&11.20&9.84&9.40&Class II? \\
3&05392933-6946147&DCL215&&19&17.3&15.5&12.82&11.26&10.38&9.62&9.83&9.57&8.90& \\
3.1&05393014-6947205&Maser&&&&&14.41&14.48&13.25&11.55&10.93&&& \\
8&05393608-6946039&DCL240?&18.62&18.06&17.73&17.29&15.54&14.4&13.50&11.38&10.83&8.96
&7.16& \\
9&05393712-6945370&&&&&&15.55&14.25&14.23&11.16&9.79&8.19&6.70&Class I \\
12&05393774-6945408&DCL287, SHV0540-6947&&&19.2&17.2&14.34&12.97&12.28&11.76&11.67&&
& \\
14&05393883-6944356&LMC X-1&&&&&13.70&13.54&13.29&13.02&12.97&&&O8 III \\
16&05394007-6944337&CPD-69 474, Sk-69-254&12.2&11.94&11.43&11.20&11.20&11.00&10.86&1
0.93&10.76&10.73&&B5Iab \\
16.1&05394002-6943521&&&&&&12.55&12.14&12.07&12.08&12.08&&& \\
16.2&05394008-6946196&DCL204/5&13.79&13.7&13.71&13.51&13.37&13.15&13.00&12.36&&&&O5+
O7Blend \\
19&05394132-6945532&DCL265&&19.62&18.52&17.18&15.13&13.92&13.44&13.17&13.23&&& \\
21&05394188-6946122&P2&&&&&15.29&14.04&12.16&9.39&8.30&7.03&5.80&Class I \\
22&05394192-6947551&&&&&&12.79&11.48&11.00&10.49&10.58&10.22&10.02& \\
23&05394222-6947010&DCL99&18.40&16.76&16.13&15.29&14.2&13.37&13.14&13.07&13.37&&& \\
24&05395040-6946436&DCL319&16.57&15.12&14.53&13.72&12.57&11.77&11.60&11.38&11.61&&&
\\
25&05394233-6945110&DCL355&18.72&17.05&16.39&15.55&14.26&13.45&13.20&12.94&13.21&&&
\\
26&05394246-6946563&DCL107&18.37&16.72&16.10&15.26&14.11&13.32&13.09&12.77&13.04&&&
\\
30&05394445-6944355&red&&&&&16.4&15.97&&14.21&13.87&11.08&9.42& \\
32&05394514-6944509&&&&&&14.90&14.44&13.95&12.75&11.84&10.04&8.06& \\
33&05394539-6948024&&&&&&16.26&14.03&13.04&11.75&11.65&11.23&10.76& \\
35&05394632-6945312&DCL304&16.39&16.13&16.04&15.75&16.02&14.85&14.52&13.84&13.86&&&b
lend \\
36&05394686-6946307&DCL168&19.10&17.65&17.07&16.24&15.11&14.23&14.12&13.85&13.97&&&
\\
48&05395181-6946325&&&&&&12.36&11.20&10.59&10.08&10.31&10.23&10.16& \\
53.1&05395279-6945175&Bubble Center, MSXG280.2128-31.4949, M7, M15&&&&&13.7&13.53&13
.35&12.99&12.83&&&blend  \\
66&05395825-6943266&&&&&&14.01&13.14&12.99&13.01&13.14&&& \\
70&05395875-6944040&R149, Sk-69 257, LMC V3936, M193&12.41&12.49&&&12.69&12.68&12.70
&12.66&12.68&&&O9III \\
72&05395881-6948165&LMC V3938, SHV0540-6949&&&&&13.38&12.03&11.15&10.54&10.35&10.16&
10.01&carbon star \\
73&05395936-6945263&P1, M210&&&&&16.46&13.97&11.84&9.39&8.68&7.76&7.61&Class I? \\
78&05400074-6947129&N159S, MSXG280.2511-31.4792&&&&&13.91&13.17&12.02&10.43&9.72&8.6
3&7.01&extended \\
84&05400177-6943465&M290&&&&&12.9&11.91&11.66&11.49&11.67&&& \\
85&05400185-6946555&&&&&&13.4&12.48&12.26&12.10&12.42&12.11&& \\
91&05400323-6942458&&&&&&12.68&11.29&10.48&9.68&9.93&10.00&9.49& \\
98&05400448-6944375&Papillion, MSXG280.1988-31.4781, M371&&&&&13.5&12.89&11.73&10.38
&9.44&7.75&6.34&extended \\
101&05400568-6945021&M410&&&&&13.67&13.13&12.73&12.27&12.27&&& \\
106&05400718-6943068&&&&&&13.08&11.66&10.93&10.24&10.23&9.96&9.51& \\
107&05400714-6943246&&&&&&12.89&12.22&12.02&11.83&11.81&11.82&& \\
121&05401355-6943506&M588&&&&&12.2&11.49&11.36&11.23&11.48&11.27&& \\
125&05401527-6946082&&&&&&13.6&12.39&12.02&11.92&12.13&12.02&12.12& \\
134&&SST 05401912-6944456&&&&&&&&14.38&13.03&11.45&10.30&Class I \\
147&05402536-6946450&KDMK 5550&&&15.2&13.73&12.72&11.59&11.03&10.40&10.52&10.25&9.72
& carbon star \\
151&05402853-6945341&&&&&&13.18&12.25&11.97&11.80&11.98&11.73&11.42& \\

\enddata
\tablerefs{DCL = \citet{deh92b}, M = \citet{mey04}, 
KDMK = \citet{kon01}, SHV = \citet{hug90} }
\end{deluxetable}

\begin{deluxetable}{cccccc}
\tablenum{3}
\tablewidth{0pt}
\tablecaption{Integrated Fluxes for HII Regions \label{aper-table}}
\tablehead{
\colhead{Object} & \colhead{Box Size} & \multicolumn{4}{c}{F$_\nu$ (mJy)} \\
\colhead{} & \colhead{$\arcsec \times \arcsec$} & \colhead{3.5~$\micron$} & \colhead{4.5~$\micron$} & \colhead{5.7~$\micron$} & \colhead{7.9~$\micron$}}
\startdata

N159AN & 26$\times$35 & 130 & 130 & 820 & 2200 \\
N159A & 40$\times$37 & 290 & 300 & 1300 & 3500 \\
Papillon & 26$\times$26 & 140 & 130 & 630 & 1900 \\
Maser & 13$\times$13 & 32 & 29 & 126 & 320 \\
N159S & 11$\times$11 & 30 & 29 & 76 & 200 \\
Center & 31$\times$27 & 140 & 140 & 520 & 1500 \\

\enddata
\end{deluxetable}

\end{document}